\documentclass[12pt]{article}
\usepackage{lscape}
\usepackage{citesort}
\usepackage{amssymb}
\usepackage{appendix}
\usepackage{multirow}

\begin{document}
\def\qq{\langle \bar q q \rangle}
\def\uu{\langle \bar u u \rangle}
\def\dd{\langle \bar d d \rangle}
\def\sp{\langle \bar s s \rangle}
\def\GG{\langle g_s^2 G^2 \rangle}
\def\Tr{\mbox{Tr}}
\def\figt#1#2#3{
        \begin{figure}
        $\left. \right.$
        \vspace*{-2cm}
        \begin{center}
        \includegraphics[width=10cm]{#1}
        \end{center}
        \vspace*{-0.2cm}
        \caption{#3}
        \label{#2}
        \end{figure}
	}
	
\def\figb#1#2#3{
        \begin{figure}
        $\left. \right.$
        \vspace*{-1cm}
        \begin{center}
        \includegraphics[width=10cm]{#1}
        \end{center}
        \vspace*{-0.2cm}
        \caption{#3}
        \label{#2}
        \end{figure}
                }

\def\ds{\displaystyle}
\def\beq{\begin{equation}}
\def\eeq{\end{equation}}
\def\bea{\begin{eqnarray}}
\def\eea{\end{eqnarray}}
\def\beeq{\begin{eqnarray}}
\def\eeeq{\end{eqnarray}}
\def\ve{\vert}
\def\vel{\left|}
\def\ver{\right|}
\def\nnb{\nonumber}
\def\ga{\left(}
\def\dr{\right)}
\def\aga{\left\{}
\def\adr{\right\}}
\def\lla{\left<}
\def\rra{\right>}
\def\rar{\rightarrow}
\def\lrar{\leftrightarrow}  
\def\nnb{\nonumber}
\def\la{\langle}
\def\ra{\rangle}
\def\ba{\begin{array}}
\def\ea{\end{array}}
\def\tr{\mbox{Tr}}
\def\ssp{{\Sigma^{*+}}}
\def\sso{{\Sigma^{*0}}}
\def\ssm{{\Sigma^{*-}}}
\def\xis0{{\Xi^{*0}}}
\def\xism{{\Xi^{*-}}}
\def\qs{\la \bar s s \ra}
\def\qu{\la \bar u u \ra}
\def\qd{\la \bar d d \ra}
\def\qq{\la \bar q q \ra}
\def\gGgG{\la g^2 G^2 \ra}
\def\GG{\langle g_s^2 G^2 \rangle}
\def\g5{\gamma_5 \not\!q}
\def\x{\gamma_5 \not\!x}
\def\g5{\gamma_5}
\def\sb{S_Q^{cf}}
\def\sd{S_d^{be}}
\def\su{S_u^{ad}}
\def\sbp{{S}_Q^{'cf}}
\def\sdp{{S}_d^{'be}}
\def\sup{{S}_u^{'ad}}
\def\ssp{{S}_s^{'??}}

\def\sig{\sigma_{\mu \nu} \gamma_5 p^\mu q^\nu}
\def\fo{f_0(\frac{s_0}{M^2})}
\def\ffi{f_1(\frac{s_0}{M^2})}
\def\fii{f_2(\frac{s_0}{M^2})}
\def\O{{\cal O}}
\def\sl{{\Sigma^0 \Lambda}}
\def\es{\!\!\! &=& \!\!\!}
\def\ap{\!\!\! &\approx& \!\!\!}
\def\ar{&+& \!\!\!}
\def\arrr{\!\!\!\! &+& \!\!\!}
\def\ek{&-& \!\!\!}
\def\vev{&\vert& \!\!\!}
\def\kek{\!\!\!\!&-& \!\!\!}
\def\cp{&\times& \!\!\!}
\def\se{\!\!\! &\simeq& \!\!\!}
\def\eqv{&\equiv& \!\!\!}
\def\kpm{&\pm& \!\!\!}
\def\kmp{&\mp& \!\!\!}
\def\mcdot{\!\cdot\!}
\def\erar{&\rightarrow&}


\def\simlt{\stackrel{<}{{}_\sim}}
\def\simgt{\stackrel{>}{{}_\sim}}


\title{
         {\Large
                 {\bf
Magnetic moment for the negative parity $\Lambda \to \Sigma^0$ transition
in light cone QCD sum rules
                 }
         }
      }

\author{\vspace{1cm}\\
{\small T. M. Aliev \thanks {e-mail:
taliev@metu.edu.tr}~\footnote{permanent address:Institute of
Physics,Baku,Azerbaijan}\,\,,
M. Savc{\i} \thanks
{e-mail: savci@metu.edu.tr}} \\
{\small Physics Department, Middle East Technical University,
06800 Ankara, Turkey }}

\date{}

\begin{titlepage}
\maketitle
\thispagestyle{empty}

\begin{abstract}

The magnetic moment of the $\Lambda \to \Sigma^0$ transition between negative parity,
baryons is calculated in framework of the QCD sum rules approach, using the
general form of the interpolating currents. The pollution arising from the
positive--to--positive, and positive to negative parity baryons are
eliminated by constructing the sum rules for different Lorentz structures.
Nonzero value of the considered magnetic moment can be attributed
to the violation of the $SU(3)$ symmetry.

\end{abstract}

~~~PACS numbers: 11.55.Hx, 13.40.Em, 14.20.Lq, 14.20.Mr,

\end{titlepage}

\section{Introduction}

Magnetic moment of baryons is one of the most important quantities in
investigation of their electromagnetic structure, and can provide essential
information about the dynamics of the strong interaction at low energies.
The magnetic moments of the octet baryons have already been calculated in
various theoretical approaches. these calculations have the privilege that
they can be checked against the available precise experimental data. The
study of the $\Lambda \to \Sigma^0$ transition magnetic moment can play
critical role in investigation of the properties of the octet baryons.

In recent years the study of the negative parity baryons have become of the
most promising direction on connection with the experiments conducted and
planned at Jefferson laboratory \cite{Rcan01}, and Mainz Microtron facility
(MAMI) \cite{Rcan02,Rcan03}. The magnetic moments of $N^\ast$ are
planned to be measured at MAMI \cite{Rcan03,Rcan04}. In the present
work we calculate the transition magnetic moment between the negative parity
$\Lambda^\ast$ and $\Sigma^{0\ast}$ baryons within the QCD sum rules method (LCSR)
(here and in further discussions, we denote the negative parity baryons as
$B^\ast$). This method is based on operator product expansion (OPE) near
light cone. The OPE is performed over the twist of the operators rather than dimension,
as is the case in the traditional QCD sum rules method. In this version all
nonperturbative dynamics is encoded in light cone distribution amplitudes.
These amplitudes appear when the matrix elements of the nonlocal operators
are sandwiched between the vacuum and one--particle states (about the
details of the LCSR see \cite{Rcan05}).
The magnetic moment of the $\Lambda \to \Sigma^0$ transition has
already been calculated in framework of the traditional QCD sum rules
\cite{Rcan06},
the external field method in the traditional QCD sum rules \cite{Rcan07},
and in the light cone version of the QCD sum rules method \cite{Rcan08}.
Note that the magnetic moments of
the negative parity octet baryons, $J^P=3^-/2$ heavy baryons, as well as diagonal
and transition magnetic moments of negative party heavy baryons are calculated
within the same framework in
\cite{Rcan09}, \cite{Rcan10} and \cite{Rcan11}, respectively.

The work is arranged as follows. In section 2 the LCSR for the magnetic
moment of the $\Lambda^\ast \to \Sigma^{0\ast}$ transition is derived. In
section 3 we numerically analyze these LCSR obtain for the transition
magnetic moment. This section also contains concluding remarks.

\section{Light cone QCD sum rules for the magnetic moment of the
$\Lambda^\ast \to \Sigma^{0\ast}$ transition}

In order to obtain the light cone sum rules for the magnetic moment of the
$\Lambda^\ast \to \Sigma^{0\ast}$ transition the following time ordered
correlation function in the vacuum in presence of the external magnetic
field is considered,
\bea
\label{ecan01}
\Pi = i \int d^4x e^{ipx} \lla 0 \vel \mbox{T} \left\{ \eta_{\Sigma^0} (x)
\bar{\eta}_\Lambda (0) \right\} \ver 0 \rra_\gamma~,
\eea
where $\eta_B$ is the interpolating current of the corresponding baryon. 
Firstly, on the phenomenological side the calculation is carried out by
saturating a tower of hadronic intermediate states carrying the same quantum
numbers as the interpolating current. Secondly, on the QCD side it is described
in terms of quarks and gluons. The QCD sum rules is constructed by matching
these two representations. The interpolating currents needed in calculation
of the correlation function  are constructed from the quark fields with the
same quantum numbers of the corresponding baryon. The general form of the
interpolating currents of $\Lambda$ and $\Sigma^0$ baryons are \cite{Rcan12}:
\bea
\label{ecan02}
\eta_\Lambda \es 2 \sqrt{1\over 6} \varepsilon^{abc} \Big\{2 (u^{aT} C    
d^b) \gamma_5 s^c + (u^{aT} C s^b)\gamma_5 d^c - (d^{aT} C s^b)\gamma_5 u^c +
2 \beta (u^{aT} C \gamma_5 d^b) s^c \nnb \\
\ar \beta (u^{aT} C \gamma_5 s^b) d^c -
\beta (d^{aT} C \gamma_5 s^b) u^c \Big\}~,\nnb \\
\eta_{\Sigma^0} \es  \sqrt{2} \varepsilon^{abc} \Big\{(u^{aT} C s^b)
\gamma_5 d^c + (d^{aT} C s^b) \gamma_5 u^c + \beta (u^{aT} C \gamma_5 s^b) d^c
+ \beta (d^{aT} C \gamma_5 s^b) u^c \Big\}~,
\eea
where $a,b,c$ are the color indices, $C$ is the charge conjugation
operator, superscript T denotes the transpose operator, and $\beta$ is the
arbitrary parameter with $\beta=-1$ corresponding to the Ioffe current.

Firstly we shall calculate the phenomenological part of the correlation
function given in Eq. (\ref{ecan01}). Saturating the interpolating current
with the intermediate hadronic states having the same quantum number as the
interpolating currents, and isolating the ground state contributions we get,
\bea
\label{ecan03}
\Pi \es {\lla 0 \vel \eta_{\Sigma^0} \ver \Sigma^0(p_2) \rra \over
p_2^2-m_{\Sigma^0}^2} \lla \Sigma^0(p_2) \gamma(q) \ve \Lambda(p_1) \rra 
{\lla \Lambda(p_1) \vel \bar{\eta}_\Lambda \ver 0 \rra \over
p_1^2-m_{\Lambda}^2} \nnb \\
\ar {\lla 0 \vel \eta_{\Sigma^{0\ast}} \ver \Sigma^{0\ast}(p_2) \rra \over
p_2^2-m_{\Sigma^{0\ast}}^2} \lla \Sigma^{0\ast}(p_2) \gamma(q) \ve
\Lambda^\ast (p_1) \rra 
{\lla \Lambda^\ast(p_1) \vel \bar{\eta}_\Lambda \ver 0 \rra \over                
p_1^2-m_{\Lambda^\ast}^2} \nnb \\
\ar {\lla 0 \vel \eta_{\Sigma^0} \ver \Sigma^0(p_2) \rra \over
p_2^2-m_{\Sigma^0}^2} \lla \Sigma^0(p_2) \gamma(q) \ve \Lambda^\ast(p_1) \rra 
{\lla \Lambda^\ast(p_1) \vel \bar{\eta}_\Lambda \ver 0 \rra \over
p_1^2-m_{\Lambda^\ast}^2} \nnb \\
\ar {\lla 0 \vel \eta_{\Sigma^{0\ast}} \ver \Sigma^{0\ast}(p_2) \rra \over
p_2^2-m_{\Sigma^{0\ast}}^2} \lla \Sigma^{0\ast}(p_2) \gamma(q) \ve
\Lambda (p_1) \rra 
{\lla \Lambda(p_1) \vel \bar{\eta}_\Lambda \ver 0 \rra \over                
p_1^2-m_{\Lambda}^2}~,
\eea
where superscript $\ast$ means it is a negative parity baryon. the matrix
elements in Eq. (\ref{ecan03}) are determined in the following way:
\bea
\label{ecan04}
\lla 0 \vel \eta \ver B(p) \rra \es \lambda_B u(p)~,\nnb \\
\lla 0 \vel \eta \ver B^\ast(p) \rra \es \lambda_{B^\ast}\gamma_5 u(p)~,\nnb \\
\lla B_2(p_2) \gamma (q)  \ve  B_1(p_1) \rra \es 
e \varepsilon^\mu \bar{u}(p_2) \Bigg[ f_1 \gamma_\mu - i {\sigma_{\mu\mu}
q^\nu \over m_{B_1} + m_{B_2}} f_2 \Bigg] u(p_1)~,\nnb \\
\lla B_2^\ast (p_2) \gamma (q)  \ve B_1^\ast (p_1) \rra \es 
e \varepsilon^\mu \bar{u}(p_2) \Bigg[ f_1^\ast \gamma_\mu - i {\sigma_{\mu\mu}
q^\nu \over m_{B_1^\ast} + m_{B_2^\ast}} f_2^\ast \Bigg] u(p_1)~,\nnb \\
\lla B_2^\ast (p_2) \gamma (q)  \ve B_1 (p_1) \rra \es 
e \varepsilon^\mu \bar{u}(p_2) \Bigg[ f_1^T \gamma_\mu - i {\sigma_{\mu\mu}
q^\nu \over m_{B_1} + m_{B_2^\ast}} f_2^T \Bigg] \gamma_5 u(p_1)~.
\eea
Substituting these matrix elements into Eq. (\ref{ecan03}), and performing
summation over the spins of the baryons we get,
\bea
\label{ecan05}
&&A^\prime \left( \not\!{p}_2 + m_{\Sigma^0}\right) \not\!{\varepsilon} \left(
\not\!{p}_1 + m_{\Lambda}\right) +
B^\prime \left( \not\!{p}_2 - m_{\Sigma^{0\ast}}\right) \not\!{\varepsilon} \left( 
\not\!{p}_1 - m_{\Lambda^\ast}\right) \nnb \\
\ar C^\prime \left( \not\!{p}_2 - m_{\Sigma^{0\ast}}\right)
\not\!{\varepsilon} \left( \not\!{p}_1 + m_{\Lambda}\right) +
D^\prime \left( \not\!{p}_2 + m_{\Sigma^0}\right) \not\!{\varepsilon} \left(
\not\!{p}_1 - m_{\Lambda^\ast}\right) + \cdots~,\nnb \\
\eea
where
\bea
\label{ecan06}
A^\prime \es { \lambda_{\Sigma^0}(\beta) \lambda_\Lambda (\beta) \over
(m_{\Sigma^0}^2 - p_2^2) (m_\Lambda^2 - p_1)^2} (f_1 + f_2)~,\nnb \\
B^\prime \es { \lambda_{\Sigma^{0\ast}} (\beta) \lambda_{\Lambda^\ast} (\beta) \over
(m_{\Sigma^{0\ast}}^2 - p_2^2) (m_{\Lambda^\ast}^2 - p_1)^2} (f_1^\ast +
f_2^\ast)~,\nnb \\
C^\prime \es { \lambda_{\Sigma^{0\ast}} (\beta) \lambda_{\Lambda} (\beta) \over
(m_{\Sigma^{0\ast}}^2 - p_2^2) (m_{\Lambda}^2 - p_1)^2} \Bigg(f_1^T +
{m_{\Sigma^{0\ast}} - m_{\Lambda} \over m_{\Sigma^{0\ast}} +
m_{\Lambda}}f_2^T\Bigg)~,\nnb \\
D^\prime \es - { \lambda_{\Sigma^0} (\beta) \lambda_{\Lambda^\ast} (\beta) \over
(m_{\Sigma^0}^2 - p_2^2) (m_{\Lambda^\ast}^2 - p_1)^2} \Bigg(f_1^T +   
{m_{\Lambda^\ast} - m_{\Sigma^0} \over m_{\Lambda^\ast} + m_{\Sigma^0}} f_2^T\Bigg)~,
\eea
where dots denote rest of the structures other that $\gamma_\mu$. The
$\Lambda^\ast \to \Sigma^{0\ast}$ transition magnetic
moment in natural units is described by $f_1^\ast+f_2^\ast$ at the point $q^2=0$.
Therefore, in order to determine the magnetic moment of the
$\Lambda^\ast \to \Sigma^{0\ast}$ transition the four equations in Eq.
(\ref{ecan05}) should be solved.

The result of the calculation for the correlation function from the QCD side
can be obtained from the diagonal $\Sigma^{0\ast}$--$\Sigma^{0\ast}$
transition as follows. It is noted in \cite{Rcan13} that the magnetic moment
for the $\Lambda$--$\Sigma^0$ transition can be determined from the diagonal
$\Sigma^0$--$\Sigma^0$ transition by using the relation between the
correlation function which is given as (more precisely using the relation
between the invariant functions for the different Lorentz structures),
\bea
\label{ecan07}   
\Pi_i^{\Sigma^0(u \lrar s)} - \Pi^{\Sigma^0(u \lrar d)} = \sqrt{3}
\Pi_i^{\Sigma^0\Lambda}~.
\eea
This relation shows that one can obtain the QCD sum rules for the
$\Lambda^\ast$--$\Sigma^{0\ast}$ transition magnetic moment by making simple
substitutions in the result for the diagonal
$\Sigma^{0\ast}$--$\Sigma^{0\ast}$ transition.

The invariant functions for the diagonal $\Sigma^{0\ast}$--$\Sigma^{0\ast}$
transition
are calculated in \cite{Rcan09} (see also the Appendix in \cite{Rcan14}), and in
the same manner with
the help of Eq. (\ref{ecan07}) the same calculation can easily be repeated
for the $\Lambda^\ast$--$\Sigma^{0\ast}$ transition. For this reason, in the present
work we do not present
the result of the correlation function from the QCD side.

As has already been mentioned, in order to determine the magnetic moment of
the negative parity $\Lambda^\ast$--$\Sigma^{0\ast}$ transition four
equations are needed. In constructing these four equations we need four
Lorentz structures. In the present work we choose the structures
$\rlap/{p}\rlap/{\varepsilon}\rlap/{q}$,
$\rlap/{p}\rlap/{\varepsilon}$, $\rlap/{\varepsilon}\rlap/{q}$,
$\rlap/{\varepsilon}$, and denote the
corresponding invariant functions as $\Pi_1$, $\Pi_2$, $\Pi_3$
and $\Pi_4$, respectively.

The sum rules for the $\Lambda^\ast$--$\Sigma^{0^\ast}$ transition is derived by
equating the coefficients of the structures 
$\rlap/{p}\rlap/{\varepsilon}\rlap/{q}$,                                    
$\rlap/{p}\rlap/{\varepsilon}$, $\rlap/{\varepsilon}\rlap/{q}$,             
$\rlap/{\varepsilon}$
of the correlation
from the from the phenomenological and QCD side, and perform double Borel
transformation over the variables $p_1^2=(p+q)^2$ and $p_2^2=p^2$, and then
solve the system of algebraic equations. As the result of these steps of
calculations we get the following expression for the magnetic moment of the
negative parity $\Lambda$--$\Sigma^0$ transition,
\bea
\label{ecan08}
\mu \es {e^{m_{\Sigma^{0\ast}}^2}/M^2 \over \lambda_{\Lambda^\ast}
\lambda_{\Sigma^{0\ast}} (m_{\Sigma^0} + m_{\Sigma^{0\ast}})
(m_{\Sigma^0}^2 + 3 m_{\Sigma^{0\ast}}^2)} \Big\{\Big[ m_{\Sigma^0} (
m_{\Sigma^0} - m_{\Sigma^{0\ast}}) - 2 m_{\Sigma^{0\ast}}^2 \Big] \Pi_1^B \nnb \\
\ek 2 m_{\Sigma^0} (m_{\Sigma^0} + m_{\Sigma^{0\ast}}) \Pi_2^B -
(m_{\Sigma^0} - 3 m_{\Sigma^{0\ast}}) \Pi_3^B -
m_{\Sigma^0} (m_{\Sigma^0} + m_{\Sigma^{0\ast}}) \Pi_4^B \Big\}~,
\eea
where we have used $M_1^2-M_2^2=2 M^2$ and $m_\Lambda \simeq m_{\Sigma^0}$,
$m_{\Lambda^\ast} \simeq m_{\Sigma^{0\ast}}$. The residues
$\lambda_{\Lambda^\ast}$ and $\lambda_{\Sigma^{0\ast}}$ are calculated in
\cite{Rcan09}.

Here, few remarks about the calculation of the correlation function from the
QCD side are in order. This correlation function contains three different
contributions. a) Perturbative part, which corresponds to the case when
photon interacts with quarks perturbatively, and all propagators of the free
quarks are considered. b) Mixed part which corresponds to the case when
photon interacts with quarks perturbatively, and at least one quark
propagator is replaced by the corresponding condensates. c) Nonperturbative
part. In this case photon interacts with the quarks at long distance. This
interaction is described by the matrix element of the nonlocal operators
between the vacuum and one--photon states, i.e.,
\bea
\label{no label04}   
\lla \Gamma (q) \vel \bar{q} \Gamma_i \left( G_{\mu\nu} \Gamma_i \right) q
\ver 0 \rra~.\nnb
\eea
These matrix elements are parametrized in terms of the photon distribution
amplitudes (DAs). The definitions of the above--mentioned matrix elements
and photon DAs are presented in \cite{Rcan15}.

\section{Numerical results}

This section is devoted to the numerical analysis of the sum rules obtained
for the $\sigma^0$--$\Lambda$ transition magnetic moment of the negative
parity baryons. The values of the input parameters entering to the sum rules
are $\uu (1~GeV) = \dd (1~GeV) = -(0.243)^3~GeV^3$,  $\sp (1~GeV) =
0.8 \uu (1~GeV)$, $m_0^2=(0.8\pm 0.2)~GeV^2$ \cite{Rcan16}, $\Lambda = (0.5
\pm 0.1)~GeV$ \cite{Rcan17}, $f_{3\gamma} = -0.039$ \cite{Rcan15}. The value
of the magnetic susceptibility is determined from the QCD sum rules analysis
to have the value $\chi (1 ~GeV) = - (2.85 \pm 0.5)~GeV^{-2}$ \cite{Rcan18},
and $m_s(2~GeV)=(111 \pm 6)~MeV$ \cite{Rcan19}. Also,
the expressions of the photon DAs, which are the
main ingredients of the LCSR, are presented in \cite{Rcan15}.  

The sum rules for the transition magnetic moment of the
$\Lambda^\ast$--$\Sigma^{0\ast}$ transition contains three auxiliary
parameters, namely, the continuum threshold $s_0$, the arbitrary parameter
$\beta$ in the interpolating current, the Borel mass parameter $M^2$;
and he magnetic moment should be independent of them.

The working region of the Borel mass parameter $M^2$ for the magnetic
moment of $\Lambda^\ast$ and $\Sigma^{0\ast}$ transition is determined in \cite{Rcan09}
to have the range $1.6~GeV^2 \le M^2 \le 3.0~GeV^2$.
Since we take  $m_\Lambda \simeq m_{\Sigma^0}$ we can also use the same domain  
in the present analysis.

The second arbitrary parameter of the sum rules is the continuum threshold
$s_0$. This parameter is related by the energy of the first state. The energy
difference between the first and ground states ranges from $0.3~GeV$ to
$0.8~GeV$. In our calculations we use the average value
$\sqrt{s_0}=(m_{ground} + 0.5)~GeV$. Finally, in order to determine the
domain of the arbitrary parameter $\beta$ that appears in the interpolating
current, we consider the dependence of the magnetic moment on $\cos\theta$,
where $\beta=\tan\theta$, at several fixed values of $M^2$ and $s_0$
chosen from their respective working regions.

As the result of our detailed numerical study, the magnetic moment of the
negative parity $\Lambda^\ast$ and $\Sigma^{0\ast}$ transition is found to
have the value,
\bea
\label{nolabel05}
\mu_{\Lambda^\ast\Sigma^{0\ast}} = (0.2 \pm 0.05) \mu_N~,\nnb
\eea
where the error in the result can be attributed to the variation in $M^2$
and $s_0$, as well as the uncertainties in the photon DAs and input
parameters.

In the $SU(3)$ limit the $\Lambda^\ast$--$\Sigma^{0\ast}$ transition
magnetic moment is equal to zero. The deviation from zero is a measure of
the $SU(3)$ symmetry violation, and we see that this violation is about 20\%.

In conclusion, the magnetic moment of the $\Lambda$--$\Sigma^0$ transition
for the negative parity baryons is estimated in framework of the LCSR.
It is obtained that the violation
of $SU(3)$ symmetry leads to nonzero value for the
$\Lambda^\ast$--$\Sigma^{0\ast}$ transition magnetic moment.

\newpage


\end{document}